\documentclass[aps,prd,reprint,groupedaddress,amssymb,amsmath]{revtex4-1}
\usepackage{bm}
\usepackage{hyperref}
\usepackage{makeidx}
\usepackage{amsmath}
\usepackage{graphicx}
\usepackage{dcolumn}
\usepackage{bm}
\usepackage{color}
\usepackage{amsfonts}
\usepackage{amssymb}

\begin{document}

\title{Quantum corrections for the phase diagram of systems with competing order}

\author{N. L. Silva J\'unior$^{1}$}
\email{nlsjunior12@gmail.com}
\author{Mucio A. Continentino$^{1}$}
\author{Daniel G. Barci$^{2}$}


\affiliation{$^{1}$Centro Brasileiro de Pesquisas F\'{\i}­sicas, Rua Dr. Xavier Sigaud 150, Urca, 22290-180, Rio de Janeiro , Brazil}
\affiliation{$^{2}$Departamento de F\'{\i}­sica Te\'orica, Universidade do Estado do Rio de Janeiro, Rua S\~ao Francisco Xavier 524, 20550-013,
Rio de Janeiro, RJ, Brazil.}

\date{\today}

\begin{abstract}
We use the effective potential method of quantum field theory to obtain the quantum corrections to the zero temperature phase diagram of systems with competing order parameters. We are particularly interested in two different scenarios: regions of the phase diagram where there is a bicritical point, at which both phases vanish continuously, and the case where both phases  coexist homogeneously.  We consider different types of couplings between the order parameters, including a bilinear one. This kind of coupling breaks time-reversal symmetry  and it is only allowed if both order parameters  transform according to the same irreducible representation. This occurs in many physical systems of actual interest like competing spin density waves, different types of orbital antiferromagnetism, elastic instabilities of crystal lattices, vortices in a multigap SC and also applies to describe the unusual  magnetism of the heavy fermion compound URu$_{2}$Si$_{2}$. Our results show that quantum corrections have an important effect on the phase diagram of systems with competing orders.
\end{abstract}

\maketitle

\section{Introduction}

The study of strongly correlated  electronic systems (SCES) constitute an exciting area of condensed matter physics. It deals with the interesting, emergent phenomena that arise from the interactions between the electrons.  Among the different types of ground states presented by SCES we find, for example,  ferromagnetism, spin density wave, Mott insulator, unconventional superconductivity, heavy Fermi liquid and charge density wave~\cite{cdw,0}. In many cases, these systems can sustain different ground states  for  the same values  of the external parameters. This leads to a competition between different orderings and eventually to coexistence.  In general SCES are sensitive to small changes of a control parameter~\cite{0,sachdev}, as external pressure, applied magnetic field or doping levels. This allows to study these materials as they transit through quantum phase transitions between the different ground states as the external parameter is varied. 

Among the SCES materials a very interesting class comprises  inter-metallic compounds containing unstable $f$-shell elements, as Ytterbium (Yb), Cerium (Ce) and Uranium (U). Since the $f$-ions are displayed on the sites of a lattice, these compounds have lattice translation invariance and ideally their resistivity should vanish as temperature approaches zero~\cite{0,1,1.2}. The physical properties of these systems is a direct result of the competition between the Kondo effect and RKKY interactions, with the former favoring the formation of a non-magnetic ground state~\cite{1.1,rkky}. At zero temperature there is a quantum critical point (QCP) separating a magnetic phase from a heavy Fermi liquid state~\cite{sachdev,1.2}. At finite temperatures, in the magnetic side of the phase diagram, there is a line of   magnetic transitions, in general antiferromagnetic,  that vanishes at the QCP~\cite{1,1.2,1.1}.  In some heavy fermion compounds, at very low temperatures, experiments have shown that they can exhibit superconductivity (SC) near or in coexistence with the antiferromagnetic  (AF)  phase close to the magnetic QCP~\cite{2}.

Competing orders, SC-AF-Structural, also appear in the iron arsenide SC~\cite{3,4}. More recently,  it was discovered that some pnictides, such as, LaFeAs(O$_{1-x}$F$_{x}$), PrFeAs(O$_{1-x}$F$_{x}$), (Sr$_{1-x}$Na$_{x}$)Fe$_{2}$As$_{2}$ and (Ba$_{1-x}$K$_{x}$)Fe$_{2}$As$_{2}$, exhibit competing AF-SC order separated by a first-order phase transition and can only coexist in phase-separated macroscopic regions of the sample~\cite{3,5,6,7}.

Thus, experimental results show  that systems with strong correlations present many interesting phenomena and  very rich phase diagrams, with competing orders and coexistence~\cite{7}. Non-Fermi liquid behavior points to the existence of underlying quantum critical points and there is also the possibility of zero temperature first order transitions~\cite{2,belitz,belitz2,belitz3}. The study of the phase diagrams of SCES with competing orders is one of  the most fundamental issues that have not yet been clarified.

Even in systems with a single type of order,  instabilities in the phase diagram can arise due to the  coupling of the order parameter to other excitations, not necessarily associated with a symmetry breaking. For example,  materials as superconductors with charged excitations can couple to the fluctuations of the electromagnetic field~\cite{1.2}. Magnetic materials can couple to elastic excitations~\cite{larkinpikin} and in metallic ferromagnets the magnetization can couple to electron-hole excitations of the Fermi liquid~\cite{belitz,belitz2,belitz3,belitz4}. In many cases the effect of these couplings is to change the order of the transition associated with the relevant order parameter and this may occur even for zero temperature~\cite{belitz,belitz2,belitz3,belitz4}. 

For the case of competing instabilities, different  types of couplings between the order parameters are treated in the  literature. The nature of these couplings is dictated by symmetry arguments. Time-reversal invariance, for example, leads to a quartic coupling in the different fields, such as,  for quadrupolar~\cite{7,8} or spin nematic orderings~\cite{9,10}. In general symmetry precludes a bilinear coupling that, however,  is allowed if the two order parameters transform according to the same irreducible representation~\cite{15,16}. These bilinear terms are then actually necessary to describe competing types of spin-density waves (SDW)~\cite{11,12}, orbital AF orders~\cite{13}, elastic instabilities of the atomic crystal lattice~\cite{14}, vortices in a multigap SC~\cite{14.1} and the unusual magnetism present in the heavy fermion compound URu$_{2}$Si$_{2}$~\cite{15}. 

In this paper, we investigate the effect of quantum corrections for the zero temperature phase diagram of systems with different types of coupling between the order parameters. We consider the case of a bicritical point where both phases vanish continuously, and that where there is a region of coexistence between the different types of ordering in the ground state.  The problem where different orderings exist {\it in close proximity, but in different regions of the phase diagram} has been studied before~\cite{2}. Our aim is to obtain the quantum corrections and verify how they modify the classical predictions. We consider the cases where the coupling between the different order parameters is a conventional quartic coupling, but also that of a bilinear interaction, whenever it is allowed by symmetry. 

We use the effective potential method of quantum field theory~\cite{2,17,18} that provides the more direct and simple way to obtain the quantum corrections to the classical action. We work in a region of parameter space close to the QCPs of the competing ground states, such that, their order parameters are small and allow for a Landau type of expansion of the free energy. For simplicity, we describe the quantum dynamics of the critical modes of the competing phases, by  propagators associated with a dynamic exponent  $z = 1$~\cite{2,Mineevz1, Hertz, Mineevz2}.

The paper is organized as follows: in Section~\ref{sec2}  we use Landau theory for a system of  coupled order parameters to classically analyze the phase diagram of the system for different  couplings. In Section~\ref{sec3} we present the  method of the effective potential which will be  used to take into account the first order quantum corrections to the phase diagram. Section~\ref{sec4}  describes our results for the  quantum effects near to the QCP's of the different orders. Finally, we discuss the consequences of the quantum corrections in Section~\ref{sec6} and reserve the Appendix~\ref{secAP} to show mathematical details.

\section{Landau Theory - Classical analysis \label{sec2}}

The simplest way to describe classically competing orders  and their phase transitions is through Landau's theory~\cite{7,10}. The order parameters are a representation of a symmetry group of the system and have the role of describing the existence or not of a respective order or symmetry breaking. The system is assumed to be near its critical points, such that, both order parameters are small and the free energy can be expanded in terms of them. The presence of a given term in the expansion of the Landau free energy is dictated by the symmetries of the problem. The equilibrium state and its properties close to the phase transitions can be obtained by minimizing the free energy considered, in our case, as a function of the two competing order parameters.   

Let us consider initially the case of two one-component real order parameters $\phi_1$ and $\phi_2$. Thus, each order parameter transforms with an irreducible representation of the $Z_2$ group.  We consider bilinear, we well as, quadratic couplings between them. The Landau free energy density of this system takes the form,
\begin{equation}
f= \frac{a_{s}}{2}\phi_{1}^{2}+\frac{u_{s}}{4}\phi_{1}^{4}+\frac{a_{m}}{2}\phi_{2}^{2}+\frac{u_{m}}{4}\phi_{2}^{4}+\frac{u_{i}}{2}\phi_{1}^{2}\phi_{2}^{2}+\gamma \phi_{1} \phi_{2}.
\label{eq: num1}
\end{equation}
For $\gamma=0$, $f$ has two independent Ising symmetries $(Z_2 \times Z_2)$, meaning that 
$f$ is invariant under the transformations $\phi_1\to -\phi_1$ and $\phi_2\to -\phi_2$, {\em independently.} However, the term $\gamma\neq 0$, breaks this symmetry to one single Ising symmetry ($Z_2 \times Z_2\to Z_2$), corresponding to change the sign of $\phi_1$ and $\phi_2$ {\em simultaneously}. 
We work at zero temperature, such that, the quantities  $a_{s}=a_{s}(P)$, $a_{m}=a_{m}(P)$ are functions of a control parameter $P$, as pressure, for example. In the absence of the bilinear coupling $\gamma=0$, these quantities vanish at the critical points for their respective order parameters, i.e.,  at critical pressures $P_{c1}$ and $P_{c2}$, such that,  $a_{s}(P_{c1})=0$ and  $a_{m}(P_{c2})=0$.   Furthermore, we take  $u_{s,m}>0$  implying that  when the couplings $u_i$ and $\gamma$ between the order parameters vanish, we have two independent second order phase  transitions at $P_{c1}$ and $P_{c2}$. When the coupling between the order parameters,  $u_{i}$ and $\gamma$ are positive we have competition between the different phases~\cite{2,7,15,dalson}.  

Minimizing the free energy density, Eq.~(\ref{eq: num1}), with respect to both order parameters, $\phi_{1}$ e $\phi_{2}$, we obtain the following expressions for these quantities,
\begin{eqnarray}
\phi_{1}^{2}&=&\frac{\left[a_{s}u_{m}-a_{m}u_{i}+\gamma(u_{m}-u_{i})\right]}{u_{i}^{2}-u_{s}u_{m}} \nonumber \\ \phi_{2}^{2}&=&\frac{\left[a_{m}u_{s}-a_{s}u_{i}+\gamma(u_{s}-u_{i})\right]}{u_{i}^{2}-u_{s}u_{m}}
\label{eq: num2}
\end{eqnarray}

Replacing these expressions  into Eq.~(\ref{eq: num1}) we can rewrite the equilibrium free energy density as,
\begin{eqnarray}
f&=&f_s+ f_m-\frac{2a_{s}a_{m}u_{i}}{4 D_{ms}} +\frac{\gamma^{2}\left(-u_{s}-u_{m}+2u_{i}\right)}{4 D_{ms}}+\nonumber 
\\
&+&\frac{\gamma} {D_{ms}}\left[a_{s}u_{m}-a_{m}u_{i}+\gamma(u_{m}-u_{i})\right]^{1/2}\times \nonumber \\
&&\;\;\;\;\;\times\left[a_{m}u_{s}-a_{s}u_{i}+\gamma(u_{s}-u_{i})\right]^{1/2}, 
\label{eq: num3}
\end{eqnarray}

where, $f_s= \frac{a_{s}^{2}u_{m}}{4 D_{ms}}$, $f_m=  \frac{a_{m}^{2}u_{s}}{4 D_{ms}}$ and   $D_{ms}=u_{i}^{2}-u_{s}u_{m}$.
If we take $a_s=a_1(P-P_{c1})$, $a_m=a_2(P_{c2}-P)$, the coefficients of the quartic terms $u_{s,m,i}$ and $\gamma$ as positive constants, the equations above describe a variety of zero temperature phase diagrams and phase transitions as a function of the control parameter $P$. We are mainly interested  in the different phases and on the nature of the transitions that occur in the presence of the couplings between order parameters. We consider first the classical case and the effect of each type of interaction separately. Next, we study the effect of quantum corrections in the phase diagrams and quantum phase transitions. 

\subsection{Quartic coupling}

Following the same procedures above, for the case of an exclusive quartic interaction, i.e., $\gamma=0$, we get 

\begin{eqnarray}
f&=&f_{m}+\frac{(a_{s}u_{m}-a_{m}u_{i})^{2}}{4u_{m}D_{sm}}= \nonumber \\ &=&f_{s}+\frac{(a_{m}u_{s}-a_{s}u_{i})^{2}}{4u_{s}D_{sm}},
\label{eq: num6}
\end{eqnarray}
where $D_{ms}=u_{i}^{2}-u_{s}u_{m}$, as before.
Notice, from Eqs.~(\ref{eq: num6}), that phase coexistence is only possible if $D_{sm}<0$, i.e., if $u_{i}<\sqrt{u_{s}u_{m}}$, so that the energy density of the two coexisting  phases is smaller than the {\it condensation energies},  $f_{m}$ and  $f_{s}$ of the individual phases.
We call attention here for a particular point in the phase diagram where the two phases vanish at the same critical value of the control parameter, i.e., for $P_{c1}=P_{c2}=P_c$. It is easy to verify from the equations above that for $a_s=a_m=0$, both order parameters vanish at this point, even for finite $u_i$ and $D_{sm}<0$. On the other hand,  this is not necessarily the case in the presence of a finite bilinear coupling $\gamma$. This can be seen from Eq.~(\ref{eq: num2}) and Eq.~(\ref{eq: num3}) and will be discussed in more detail below.

\subsection{Bilinear coupling}

We now study the case that $u_{i} = 0$ in Eq.~(\ref{eq: num1}), but there is a finite bilinear coupling between the order parameters. The equilibrium values of the order parameters is given by
\begin{figure}[!h]
\begin{center}
\includegraphics[scale=0.8]{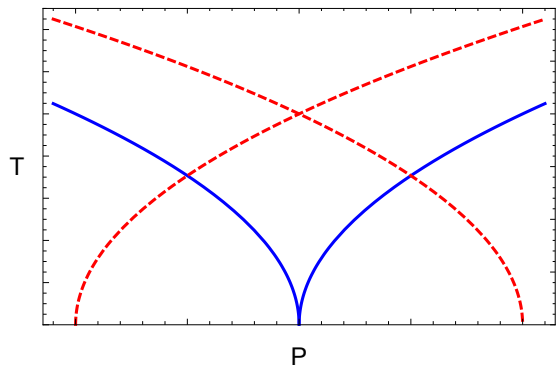} 
\end{center}
\caption{(Color online) Schematic phase diagram for both kind of couplings for $a_s=a_m=0$ and $u_{s,m}>0$. Full lines (blue) for $u_i>0$ and $\gamma=0$.  Dashed lines (red) for $u_i=0$ and $\gamma<0$. In the latter case, the bilinear coupling $\gamma$ induces a region of coexistence where both order parameters are finite at $a_s=a_m=0$  (see text). }
\label{Fig.1}
\end{figure}
\begin{eqnarray}
\phi_{1}^{2}&=&-\frac{(a_{s}+\gamma)}{u_{s}}, \nonumber \\ 
\phi_{2}^{2}&=&-\frac{(a_{m}+\gamma)}{u_{m}}
\label{eq: num8}
\end{eqnarray}
which, when substituted in the expression for the free energy density yield,
\begin{equation}
f_{c}= f_s + f_m +\gamma \sqrt{\frac{(a_{s}+\gamma)(a_{m}+\gamma)}{u_{s}u_{m}}}+\frac{\gamma^{2}}{4}\left(\frac{u_{s}+u_{m}}{{u_{s}u_{m}}}\right).
\label{eq: num9}
\end{equation}

Eq.~(\ref{eq: num9}) shows that phase coexistence is only possible for $\gamma<0$,  such that, the total free energy density of the system $f_c$ is less than the sum  $f_s+f_m$. This condition is also implicit in Eqs.~(\ref{eq: num8}).

Considering again the special point in the phase diagram,  $a_{s}=a_{m}=0$, we notice a quite different behavior from that of the previous case. Indeed from Eqs.~(\ref{eq: num8}), we realize that  at the classical level the system can have  finite order parameters induced by the bilinear coupling $\gamma$, even for $a_{s}=a_{m}=0$  (see Fig.~\ref{Fig.1}).  The phase diagrams shown in Fig.~\ref{Fig.1} are strongly affected by quantum fluctuations as will be discussed  further on in the text.  

\subsection{Both couplings present}

In this case, we can see from  Eq.~(\ref{eq: num3}) that the condition for coexistence requires that $\gamma<0$ and that $u_{i}< \sqrt{u_su_m}$.
For the special case $a_{s}=a_{m}=0$, we notice from Eq.~(\ref{eq: num2}) that both order parameters are finite due to the bilinear coupling $\gamma$ as long as $\gamma$ remains negative and the conditions  $u_m> u_i$ and $u_s > u_i$ are satisfied. Notice that the latter automatically imply  $u_{i}< \sqrt{u_su_m}$.

In the next section we are going to calculate the quantum corrections to the ground state energy of the system described by Eq.~(\ref{eq: num1}) in the \textit{one-loop approximation}. It will turn out that in this approximation, the contribution from the bilinear term to these corrections is an infinite constant independent of the fields. This problem can be circumvented applying a linear transformation with $\det=1$ (rotation) to the classical fields,
\begin{eqnarray}
\phi_{1}&=&\alpha \varphi_{1} + \beta \varphi_{2} \nonumber \\ \phi_{2}&=&-\beta \varphi_{1} +\alpha \varphi_{2}
\label{eq: num110}
\end{eqnarray}
with $\alpha^{2}+\beta^{2} = 1$. We can parametrize this transformation with just one parameter $\theta$ taking,
\begin{equation}
\alpha = \cos \theta \ \ \ \ \ ; \ \ \ \ \ \beta = \sin \theta
\label{eq: num120}
\end{equation}
Replacing  Eqs.~(\ref{eq: num110}) into the expression of the ground state energy, we get to lowest order in the fields,
\begin{equation}
f= r_{1}\varphi_{1}^{2}+r_{2}\varphi_{2}^{2}+\lambda_{1}\varphi_{1}^{4}+\lambda_{2}\varphi_{2}^{4}+\lambda_{12} \varphi_{1}^{2} \varphi_{2}^{2}+\delta_{1} \varphi_{1}^{3}\varphi_{2}+\delta_{2} \varphi_{1}\varphi_{2}^{3}
\label{eq: num130}
\end{equation}
where,
\begin{eqnarray}
r_{1}&=&\frac{a_{s}}{2}\cos^{2} \theta+\frac{a_{m}}{2}\sin^{2} \theta - \gamma \cos \theta \sin \theta  \nonumber \\ r_{2}&=&\frac{a_{s}}{2}\sin^{2} \theta +\frac{a_{m}}{2}\cos^{2} \theta +\gamma \cos \theta \sin \theta
\label{eq: num131}
\end{eqnarray}
and $\theta$  is such that,
\begin{equation}
\tan2\theta=\frac{2\gamma}{a_{m}-a_{s}}.
\label{eq: num160}
\end{equation}
The $\lambda_i$ and $\delta_i$ are new arbitrary constants.
Notice that the bilinear coupling has been replaced by new terms of higher order in the fields (with coefficients $\delta_i$), but with the same symmetry properties. The analysis of the  classical  ground state energy in this new basis is similar to that developed  before. It is important to notice that the main effect of the coupling $\gamma$ at this level  is to shift the quantum critical points (QCP),  as can be seen in Eqs.~\ref{eq: num8} and is shown in  Fig.~\ref{Fig.1}. The analysis is more simple if carried out close to the new QCPs at $r_1=0$ and $r_2=0$.

\section{Quantum corrections \label{sec3}}

The  method of the effective potential provides the most simple and direct approach to obtain information about the effects of quantum corrections  on classical phase diagrams. 
At the {\it one-loop} level, the effective potential $V_{eff}$ is given by the expression
\begin{equation}
V_{eff}(\varphi_i)=V_{cl}(\varphi_i)+\hbar\;  \Gamma^{(1)}(\varphi_i) + O(\hbar ^2)\; , 
\label{eq:Veff}
\end{equation}
where $V_{cl}(\varphi_i)$ is the classical potential (the Landau free energy density described in the previous section), and $\Gamma^{(1)}(\varphi_i)$ codifies the quantum fluctuations at linear order in  $\hbar$. $\varphi_i$ are the {\em homogeneous} order parameters. Then, the actual phases of the system are  reached by computing $\partial V_{eff}/\partial\varphi_i=0$.

Following references~\cite{2,17,18}, the quantum correction is given by
\begin{equation}
\Gamma^{(1)}(\varphi_i)=\frac{1}{2}\int \frac{d^{4}k}{(2\pi)^{4}} \ln \left(\det [1-M(k)]\right)+ \mbox{counterterms}.
\label{eq: num11}
\end{equation}
The ``counterterms'' are necessary to renormalize the theory, and $M(k)$ is a matrix whose  elements $[M]_{lm}$ are given by
\begin{equation}
[M]_{lm}=-G_{0}^{(l)}(k)\left[\frac{\partial^{2}V^{I}_{cl}({{\varphi_{i}}})}{\partial \varphi_{l}\partial \varphi_{m}}\right]_{{\varphi_{i}=\varphi_{c}}}
\label{eq: num10}
\end{equation}
Here, $G_{0}^{l}(k)$ are the propagators of the $\varphi_l$ fields and $V^{I}_{cl}$ is the interaction part of the classical potential ({\em i.\ e.\ }, the Landau  free energy density without the mass terms ($r_{1,2}$)). Dynamical effects are included through the frequency dependent propagators.

For simplicity, we will consider that the two phases have the same dynamics described by propagators associated with a dynamical exponent $z = 1$~\cite{2,Mineevz1, Hertz, Mineevz2}, that characterizes a non-dissipative behavior. It results from the relation between the nth-order time derivatives and the gradient terms in nonstationary Ginzburg-Landau equation, i.e., the frequency and wave-vector dependence of the propagators~\cite{2,Mineevz1, Hertz, Mineevz2}. However, we may have different behaviors for the propagators and consequently distinct dynamic critical exponent z. For example, $z=2$ is generally associated with a dissipative behavior~\cite{2,Mineevz2,Hertz}, among others~\cite{Hertz}. Preliminary results indicate that the choice of different dynamics, and of the dimensionality of the system can affect the final results~\cite{futur}.

The type of dynamics that we consider in this work is appropriate to magnetic phases with excitations with linear dispersion relations, superconductors~\cite{1,1.2,2, Hertz} or to superfluid liquid $^{3}$He~\cite{Mineevz1}. The relevant  propagators are given by, 
\begin{equation}
G^{(1,2)}(k)=G_{0}^{(1,2)}(\omega,\vec{q})=\frac{1}{k^{2}+r_{1,2}},
\label{eq: num12}
\end{equation}
where $k^{2}=\omega^{2}+q^{2}$ (Euclidean space).

\section{Effects of quantum corrections - Results \label{sec4}}

After the  calculation of the derivatives in Eq.~(\ref{eq: num10}) we get the expression for the first order quantum corrections for the effective potential, Eq.~(\ref{eq: num11}), as (see Appendix~\ref{secAP})
\begin{eqnarray}
\Gamma^{(1)}\!&=&\!\frac{1}{2} \!\int\!\!\!\frac{d^{4}k}{(2\pi)^{4}}\! \ln\!\left[\!\!\left(1+\frac{\left[12\lambda_{1}\varphi_{1}^{2}+2\lambda_{12}\varphi_{2}^{2}+6\delta_{1}\varphi_{1}\varphi_{2}\right]}{k^{2}+r_{1}}\right)\! \right. \nonumber \\
&\times& \left(1+\frac{\left[12\lambda_{2}\varphi_{2}^{2}+2\lambda_{12}\varphi_{1}^{2}+6\delta_{2}\varphi_{1}\varphi_{2}\right]}{k^{2}+r_{2}}\right)\!+ \nonumber  \\ 
&-&\left.\left(\frac{\left(3\delta_{1}\varphi_{1}^{2}+3\delta_{2}\varphi_{2}^{2}+4\lambda_{12}\varphi_{1}\varphi_{2}\right)^{2}}{(k^{2}+r_{1})(k^{2}+r_{2})}\right)\right]+ ct,
\label{eq: num13}
\end{eqnarray}
where {\it ct} stands for {\it counterterms}.
In order to calculate the integrals in Eq.~(\ref{eq: num13}), we perform an expansion of the  logarithm up to  second order terms in $\left( 3 \delta_{1} \varphi_{1}^{2}+3\delta_{2} \varphi_{2}^{2} + 4 \lambda_{12} \varphi_{1} \varphi_{2} \right)$ and get,
\begin{eqnarray}
\Gamma^{(1)}&=&\frac{1}{2}\!\int\!\!\!\frac{d^{4}k}{(2\pi)^{4}}\! \ln\left[\left(1+\frac{b_{1}}{k^{2}+r_{1}}\right)\left(1+\frac{b_{2}}{k^{2}+r_{2}}\right)\right] +  \nonumber \\ 
&-&\frac{\left(3\delta_{1}\varphi_{1}^{2}+3\delta_{2}\varphi_{2}^{2}+4\lambda_{12}\varphi_{1}\varphi_{2}\right)^{2}}{2(B^{2}-A^{2})} \times \nonumber \\ 
&\times&\!\int\!\!\!\frac{d^{4}k}{(2\pi)^{4}}\! \left(\frac{1}{k^{2}+A^{2}}-\frac{1}{k^{2}+B^{2}}\right) + ct,
\label{eq: num14}
\end{eqnarray}
where,
\begin{eqnarray}
b_{1}&=&12\lambda_{1}\varphi_{1}^{2}+2\lambda_{12}\varphi_{2}^{2}+6\delta_{1}\varphi_{1}\varphi_{2} ; \ A^{2}= r_{1}+b_{1} \nonumber \\
b_{2}&=&12\lambda_{2}\varphi_{2}^{2}+2\lambda_{12}\varphi_{1}^{2}+6\delta_{2}\varphi_{1}\varphi_{2} ; \ B^{2}= r_{2}+b_{2}
\label{eq: num15}
\end{eqnarray}

We can calculate the integrals in Eq.~(\ref{eq: num14}) noticing that $d^{d}k=S_{d}k^{d-1}dk$, where $S_{d}=(2 \pi)^{d/2}/\Gamma(d/2)$ and $S_{4}=2 \pi^{2}$. Therefore, the integral in the first line of Eq.~(\ref{eq: num14})~\cite{1.2,2} and the integral in the second line of Eq.~(\ref{eq: num14}) can be easily calculated (see Appendix~\ref{integrals}).

Since we are interested in the regime where both   $r_{1}$ or $r_{2}$ are  small, i.e., near to the QCP of both phases,  we will use an expansion of the effective potential in terms  of $r_{1}$ and $r_{2}$. Let us consider some particular cases.

\subsection{Bicritical point, $r_1=r_2=0$}

We start calculating the quantum corrections to the classical phase diagram when both QCPs coincide (Full lines (blue) in Fig.~\ref{Fig.1}), such that, $r_{1}=r_{2}=0$. The quantity $\Gamma^{(1)}$ for $r_{1}=r_{2}=0$ can be obtained from the previous calculations, i.e., summing Eqs.~(\ref{eq: num16.1}) and~(\ref{eq: num17.1}) (see Appendix~\ref{integrals}) and assuming $\Lambda >> 1$. We get, 
\begin{eqnarray}
\Gamma^{(1)}&=&\frac{\pi^{2}}{(2 \pi)^{4}} \bigg[ \frac{\Lambda^{2}}{2}(b_{1}+b_{2})+\frac{b_{1}^{2}}{4}\ln\left(\frac{b_{1}}{\Lambda^{2}}\right)+\frac{b_{2}^{2}}{4} \ln \left( \frac{b_{2}}{\Lambda^{2}} \right) \nonumber \\
&-&\frac{(b_{1}^{2}+b_{2}^{2})}{8} +\frac{\left(3\delta_{1}\varphi_{1}^{2}+3\delta_{2}\varphi_{2}^{2}+4\lambda_{12}\varphi_{1}\varphi_{2}\right)^{2}}{4}\times \nonumber \\
&\times&\left(\ln\left(\frac{1}{\Lambda^{2}}\right)+\frac{b_{1}\ln(b_{1})-b_{2}\ln(b_{2})}{b_{1}-b_{2}}\right) \bigg] +ct.
\label{eq: num18}
\end{eqnarray}

Proceeding as usual with the  renormalization process (see Appendix~\ref{secA}), we determine the \textit{counterterms}, with the condition that they eliminate the dependence of the effective potential on the \textit{cut-off} $\Lambda$. 
Using Eq.~(\ref{eq:Veff}), where from now on we put $\hbar=1$, we finally get,
\begin{eqnarray}
V_{eff}(\varphi_{1},\varphi_{2})&=&\tilde{\lambda}_{1}\varphi_{1}^{4}+\tilde{\lambda}_{2}\varphi_{2}^{4}+\tilde{\lambda}_{12} \varphi_{1}^{2} \varphi_{2}^{2}+\tilde{\delta}_{1} \varphi_{1}^{3}\varphi_{2}+ \nonumber \\
&+&\tilde{\delta}_{2} \varphi_{1} \varphi_{2}^{3}  + O(\varphi^{6}).
\label{eq: num19}
\end{eqnarray}
The {\it tilde} quantities represent effective couplings renormalized by quantum corrections and are given together with the terms of higher orders in the fields $O(\varphi^{6})$ in the Appendix~\ref{secB}.  Analogously to the analysis of the classical case, let us consider the quantum corrections for  some particular cases of Eq.~(\ref{eq: num19}).

\begin{figure}[!h]
\begin{center}
\includegraphics[scale=0.68]{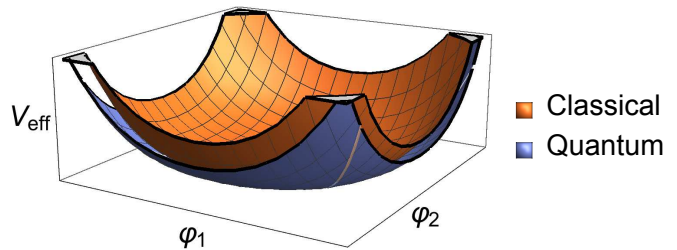} 
\end{center}
\caption{(Color online) Effective potential for a bicritical point in the case of an exclusive quartic coupling, with and without quantum corrections (schematic). There is no qualitative change in the phase diagram, i.e., the minimum of the effective potential is at the origin in both cases.}
\label{Fig.2}
\end{figure}

For an exclusive quartic coupling, i.e., taking $\delta_{i}=0$ and finite (positive) $\lambda_{12}$ on Eq.~(\ref{eq: num19}), we can verify that the first order quantum corrections do not introduce any qualitative change in the phase diagram. In other words, even taking into account quantum corrections, the minimum of the effective potential remains at the origin as can be seen in Fig.~\ref{Fig.2}. This  implies that the bicritical point is stable in the presence of quantum corrections due to a quartic coupling.  Summarizing,  quantum corrections due to an exclusive quartic coupling do not affect the phase diagram in Fig.~\ref{Fig.1} (full lines (blue line)) and the bicritical point survives these interactions at the classical and quantum level.

Next, we are interested in the case that $\lambda_{12} = 0$ and a finite (positive) bilinear coupling is allowed. For simplicity,  we can, without loss of generality,  take the limit when $\delta_{1}$ tends to $\delta_{2}$ and $\lambda_{2}$ tends to $\lambda_{1}$. In other words, the terms of $O(\varphi^{6})$ in Eq.~(\ref{eq: num19}) can be neglected. Notice also that in the presence of the bilinear coupling, the physical region of the phase diagram  is that  where the fields have the same sign, such that, time reversal symmetry is preserved. Thus, in order to verify the behavior of the system in this region of parameters, we  make a cut in the $3$D phase diagram  taking the fields $\varphi_{1}=\varphi_{2}$ in Eq.~(\ref{eq: num19}).

\begin{figure}[!h]
\begin{center}
\includegraphics[scale=0.7]{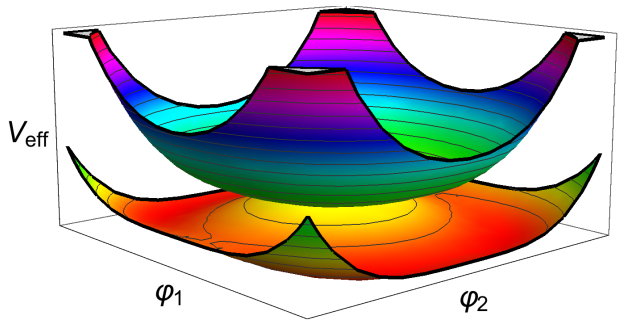} 
\end{center}
\caption{(Color online) The effective potential for a bicritical point in the case of a finite  bilinear coupling, with and without quantum corrections (schematic). When we take into account quantum corrections the minima of $V_{eff}$ move outside the origin, which means that the quantum corrections will induce coexistence in this case, even for $\delta_{i}$ positive.}
\label{Fig.3}
\end{figure}

After all these simplifications Eq.~(\ref{eq: num19}) can be rewritten in the form,
\begin{eqnarray}
&&   \;\;\;\; V_{eff}=\frac{1}{64 \pi^{2}}\varphi_{1}^{4}\bigg[\underbrace{128 \pi^{2}(\lambda_{1}+\delta_{2})}_{classical \ term} + \nonumber \\
&&   \;\;\;\;\;\;\;\;\;\;\;\; +18\delta_{2}^{2}\bigg[\ln\left(\frac{6(\varphi_{1}^{2}(2\lambda_{1}+\delta_{2}))^{2}}{\delta_{2}\left\langle \varphi\right\rangle^{2}}\right)+ \nonumber \\
&+&4\ln\left(\frac{\varphi_{1}^{2}(2\lambda_{1}+\delta_{2})}{\delta_{2}\left\langle \varphi\right\rangle^{2}}\right)-\frac{85}{6}\bigg]+288\lambda_{1}(\lambda_{1}+\delta_{2}) \times \nonumber \\
&&   \;\;\;\;\;\;\;\;\;\;\;\; \times\left[\ln\left(\frac{\varphi_{1}^{2}(2\lambda_{1}+\delta_{2})}{2\lambda_{1}\left\langle \varphi\right\rangle^{2}}\right)-\frac{25}{6}\right]\bigg]
\label{eq: num19.2}
\end{eqnarray}
where $\left\langle \varphi \right\rangle$ correspond to the minima of the effective potential (see Appendix~\ref{secA}).
We can identify from Eq.~(\ref{eq: num19.2}) two {\it Coleman-Weinberg-like} terms that give rise to minima in  the effective potential, outside the origin~\cite{18}. The plot of the $3$D effective potential in Fig.~\ref{Fig.3} shows that quantum corrections due to an exclusive bilinear coupling has a radically different effect from that of purely quartic interactions. The quantum corrections due to the former break the symmetry of the fields and induce coexistence between these orders {\it even for $\delta_{i}$ positive}.

For completeness we consider now the stability of the bicritical point when both couplings are present. Due to the presence of the bilinear coupling we can use the same arguments of the exclusive bilinear case to simplify the effective potential given in Eq.~(\ref{eq: num19}). Therefore, following the same previous procedures for the exclusive bilinear coupling, but now for $\lambda_{12}$ finite (positive) we get,
\begin{eqnarray}
&&  \;\;\;\; V_{eff}=\frac{1}{192 \pi^{2}}\varphi_{1}^{4}\bigg\{\underbrace{384 \pi^{2}(\lambda_{1}+\delta_{2}+\frac{\lambda_{12}}{2})}_{classical \ term}  \nonumber \\
&&  \;\;\;\; +54\delta_{2}^{2}\bigg[ \ln\left(\frac{(2\varphi_{1}^{2}(6\lambda_{1}+\lambda_{12}+3\delta_{2}))^{2}}{6\delta_{2}\left\langle \varphi\right\rangle^{2}}\right) + \nonumber \\
&&  \;\;\;\; +4\ln \left( \frac{\varphi_{1}^{2}(6\lambda_{1}+\lambda_{12}+3\delta_{2})}{3\delta_{2}\left\langle \varphi \right\rangle^{2}} \right) - \frac{85}{6}\bigg] 
\nonumber \\
&+& 864\lambda_{1}(\lambda_{1}+\delta_{2})\left[\ln\left(\frac{\varphi_{1}^{2}(6\lambda_{1}+\lambda_{12}+3\delta_{2})}{6\lambda_{1}\left\langle \varphi\right\rangle^{2}}\right)-\frac{25}{6}\right] \nonumber \\
&+& 24\lambda_{12}^{2}\left[\ln\left(\frac{4\left(\varphi_{1}^{2}(6\lambda_{1}+\lambda_{12}+3\delta_{2})\right)^{3}}{\lambda_{12}\left\langle \varphi\right\rangle^{2}}\right)-\frac{13}{6}\right]\nonumber \\
&+& 144\lambda_{12}\delta_{2}\left[\ln\left(\frac{4\left(\varphi_{1}^{2}(6\lambda_{1}+\lambda_{12}+3\delta_{2})\right)^{3}}{\lambda_{12}\left\langle \varphi\right\rangle^{2}}\right)-\frac{5}{2}\right] \nonumber \\ 
&+& 288\lambda_{12}\lambda_{1}\left[\ln\left(\frac{\varphi_{1}^{2}(6\lambda_{1}+\lambda_{12}+3\delta_{2})}{6\lambda_{12}\left\langle \varphi\right\rangle^{2}}\right)-\frac{7}{2}\right]\bigg\}.
\label{eq: num19.3}
\end{eqnarray}

Notice that when $\lambda_{12}=0$,  Eq.~(\ref{eq: num19.3}) above reduces to the previous studied one,  Eq.~(\ref{eq: num19.2}). In the general case, when both couplings are present, the plot of the effective potential has a behavior similar to that of the purely bilinear coupling, with minima of the effective potential outside the origin as in Fig.~\ref{Fig.3}. Then even in the presence of the quartic interaction, any bilinear coupling, $\delta_{i} > 0$ breaks the symmetry of the bicritical point and leads to a coexistence of phases for the value of parameters for which both systems were critical in the absence of $\delta_{i}$. This is confirmed by minimizing the effective potential, Eq.~\ref{eq: num19.3}, to obtain the classical field $\left\langle \varphi \right\rangle$ and expanding for small $\delta_i$ and $\lambda_{12}$. We obtain that $\left\langle \varphi \right\rangle \propto \delta^{1/2}(1 + O(\lambda_{12}^{2})+\cdots)$ implying that any finite $\delta$ gives rise to a symmetry breaking even in the presence of the quartic interaction $\lambda_{12}$.
The two couplings however are in competition, as we can easily see plotting the effective potential.
For $\lambda_{12} \ne 0$,  the plot for Eq.~(\ref{eq: num19.3}) is very similar to that of Fig.~\ref{Fig.3} with the only difference that the minima of the effective potential  occur for smaller values of the symmetry breaking field. The effect of the coupling $\lambda_{12}$ then  is to decrease the coexistence region that the $\delta_{i}$ coupling produces.

\subsection{Coexistence phase diagram}
\begin{figure}[!h]
\begin{center}
\includegraphics[scale=0.4]{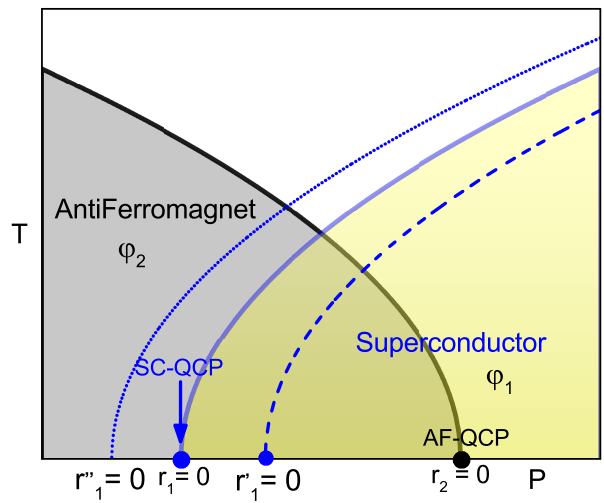} 
\end{center}
\caption{ (Color online) Schematic phase diagram of coexisting phases in the presence of both couplings and their particular cases. The full (blue) line represents the classical result and the dashed line the effect of the quantum corrections (see text). In this figure the phase with $\varphi_1\ne 0$ is identified with a superconducting phase and that with $\varphi_2 \ne 0$ with an antiferromagnet. The QCP of these phases are labelled  as SC-QCP and AF-QCP, respectively.}
\label{Fig.4}
\end{figure}

Let us consider now the more general case where both $r_1$ and $r_2$ are different from zero. Fig.~\ref{Fig.4} shows a possible schematic phase diagram for the particular case of a superconductor and an antiferromagnet where both phases coexist in a region of the phase diagram.
Classically, this occurs whenever $r_1$ and $r_2$ are negative, even in the absence of a coupling between these phases. In the presence of the quartic coupling $\lambda_{12}$ and for $\delta_1=\delta_2=0$, the required condition for coexistence, in the case of positive $\lambda_{12}$, $r_1<0$ and $r_2<0$, is that $\lambda_{12}^2<4\lambda_1 \lambda_2$. This guarantees that the coexistence of phases lowers the energy of the system. For $\lambda_{12}=0$ and $\delta_{1,2} <0$, we can check from Eq.~(\ref{eq: num130}) that  coexistence  is always possible. In the general case that $\lambda_{12}$ and $\delta_{1,2}$ are finite, we find that these quantities compete, the former favoring coexistence between phases.  The condition in this case is that $\lambda_{12}^2<4\lambda_1 \lambda_2+ \delta_2 \lambda_1 + \delta_1 \lambda_2 +\delta_1 \delta_2 /4$, with $\delta_{1,2} <0$. When these are zero we recover the previous condition for  biquadratic interactions only.

We now obtain the quantum corrections  for this case of coexistence of phases. The problem where $r_{1} \neq 0$ and $r_2 \neq 0$ is mathematically more intricate than that of the previous section.  However, we can still make analytical progress whenever the system is {\it deep} in one of the phases, but at the QCP of the other, as shown in Fig.~\ref{Fig.4}.  
Then, we treat here the problem where the material is in an  ordered phase, such that, say $r_2 < 0$ and $\varphi_2 \neq 0$,  but at the QCP of the other, i.e.,  with $r_1=0$, $\varphi_1$ small, although allowed to be finite. The symmetric case corresponding to $r_1$ small and negative, such that, the system is in the ordered $\varphi_1$ phase, but  is tuned to the  QCP of $\varphi_2$, i.e., to  $r_2=0$ can be treated in the same way and yields equivalent results since the dynamics of the propagators are considered identical in this paper. 
For the former conditions, the effective potential is obtained expanding $\Gamma_{\varphi_{1},\varphi_{2}}^{(1)}$ for $r_2$ small but $r_1=0$. We get,   
\begin{eqnarray}
\Gamma^{(1)}&=&\frac{\pi^{2}}{(2 \pi)^{4}}\bigg[\frac{\Lambda^{2}}{2}(b_{1}+b_{2})+\frac{b_{1}^{2}}{4} \ln\left( \frac{b_{1}}{\Lambda^{2}} \right) + \nonumber \\
&-&\frac{(b_{1}^{2}\!+\!b_{2}^{2})}{8} \!+\! \frac{(r_{2}+b_{2})^{2}}{4}\ln\left(\!\frac{r_{2}+b_{2}}{\Lambda^{2}}\!\right)-\frac{r_{2}^{2}}{4}\ln\left(\frac{r_{2}}{\Lambda^{2}}\right) \!+\! \nonumber \\
&+&\frac{\left(3\delta_{1}\varphi_{1}^{2}\!+\!3\delta_{2}\varphi_{2}^{2}\!+\!4\lambda_{12}\varphi_{1}\varphi_{2}\right)^{2}}{4}\times  \nonumber \\
&\times& \left(\ln\left(\frac{1}{\Lambda^{2}}\right)+\frac{B^{2}\ln(B^{2})-b_{1}\ln(b_{1})}{B^{2}-b_{1}}\right) \bigg] + \ ct.
\label{eq: num32}
\end{eqnarray}
where $B^{2}=r_{2}+b_{2}$, $b_{1}$ and $b_{2}$ are given in Eq.~(\ref{eq: num15}). Notice that both fields $\varphi_1$ and $\varphi_2$ are kept finite in this expansion even taking $r_{1} = 0$ because of the possibility of $\varphi_1$ being induced by the couplings. 

Proceeding with the renormalization process to eliminate the \textit{cut-off}, we obtain the \textit{counterterms} and the following expression for the effective potential,
\begin{eqnarray}
V_{eff}(\varphi_{1},\varphi_{2})\!&=&\! r_{2}\varphi_{2}^{2}\!+\!\lambda_{1}'\varphi_{1}^{4}\!+\!\lambda_{2}'\varphi_{2}^{4}\!+\!\lambda_{12}' \varphi_{1}^{2} \varphi_{2}^{2}+ \nonumber \\
&+&\delta_{1}'\varphi_{1}^{3} \varphi_{2}+\delta_{2}'\varphi_{1} \varphi_{2}^{3} +\tilde{\rho}(\varphi_{1,2}) +\nonumber \\
&+&\frac{1}{(4 \pi)^{2}} \eta  \left[\ln\left(\frac{B^{2}}{r_{2}}\right)-\frac{1}{2}\right]
\label{eq: num35}
\end{eqnarray}
where $\tilde{\rho}(\varphi_{1,2})$ contains terms of higher orders in the fields and,
\begin{equation}
\eta = \left(3r_{2}\delta_{2}\varphi_{1}\varphi_{2}+6r_{2}\lambda_{2}\varphi_{2}^{2}+r_{2}\lambda_{12}\varphi_{1}^{2}\right).
\label{eq: num36}
\end{equation}

Again the \textit{prime} quantities represent effective couplings renormalized by quantum corrections and are given together with the higher orders terms $\tilde{\rho}(\varphi_{1,2})$ in the Appendix~\ref{secC}. The terms $B^{2}=r_{2}+b_{2}$, $b_{1}$ and $b_{2}$, are given by Eqs.~(\ref{eq: num15}) above.

The effective potential, Eq.~(\ref{eq: num35}), specifically in the $\eta$ contribution,  contains  a $\varphi_{2}^{2}$ term with a small coefficient whose effect is just to renormalize the classical potential and does not produce qualitative changes in the classical phase diagram. However, the presence of a $\varphi_{1}^{2}$ term in $\eta$ gives rise to interesting physical consequences that we now  analyze  in detail.

First  notice that for $\delta_i = 0$  in the  effective potential, Eq.~(\ref{eq: num35}), the term in brackets [$ \cdots$] multiplied by $\eta$ and involving $\varphi_{1}^{2}$ is always negative. This term in turn couples to the product  $r_{2}\lambda_{12}$ that is also negative since the system is in the ordered $\varphi_2$ phase, i.e., $r_{2}<0$. This implies that the  coefficient of the $\varphi_{1}^{2}$ term due to the quantum correction, $\approx -1/(4 \pi)^2 r_2 \lambda_{12}$,   is always positive. The physical significance of this positive $\varphi_{1}^{2}$ term is that  the QCP of the $\varphi_1$ phase has been pushed away towards the QCP of the $\varphi_2$ phase due to the competition introduced by $\lambda_{12}$ coupling between these phases. The system that was at the QCP of the $\varphi_1$ phase is now in its disordered phase.  The deeper the system is in the $\varphi_2$ phase, the larger is this effect since $|r_{2}|$ is larger. The  effect also increases with  the intensity of the interaction $\lambda_{12}$ clearly manifesting the competitive nature of this coupling that acts in the sense of reducing the region of  coexistence in the phase diagram.  

If the same analysis is carried out at the new QCP, $r^{\prime}_1$ (see Fig.~\ref{Fig.4}),  the same effect occurs until $r^{\prime}_1=r^{\prime}_2$ and we arrive at the stable situation studied previously of a quantum bicritical point. For completeness, we point out the rather trivial case of a negative $\lambda_{12}$ in which case the opposite effect is observed and coexistence is enhanced by this coupling. As shown in Fig.~\ref{Fig.4} the new QCP of the $\varphi_1$ phase has moved to $r^{\prime \prime}_1$ due to the negative interaction.

Next we consider the case the quartic interaction $\lambda_{12}=0$, but we turn on the  couplings $\delta_{1,2}$. The terms multiplied by $\eta$ that arise in the quantum corrections due to these couplings are proportional to $r_{2}\delta_{2}$ and in particular there are no terms in $\varphi_1^2$ coupled to $\delta_{1,2}$  as can be seen from Eq.~(\ref{eq: num35}). However, the coefficient of the term $\varphi_1\varphi_2$ has an opposite behavior to that obtained for the coupling $\lambda_{12}$, since both $r_{2}$ and $\delta_{2}$ are negative, with the negative sign coming from the terms in brackets of  Eq.~(\ref{eq: num35}). We can then state that the quantum corrections arising from these couplings unlike the case of the previous coupling $\lambda_{12}$ favor an increase of the region of coexistence. 

\begin{figure}[!h]
\begin{center}
\includegraphics[scale=0.35]{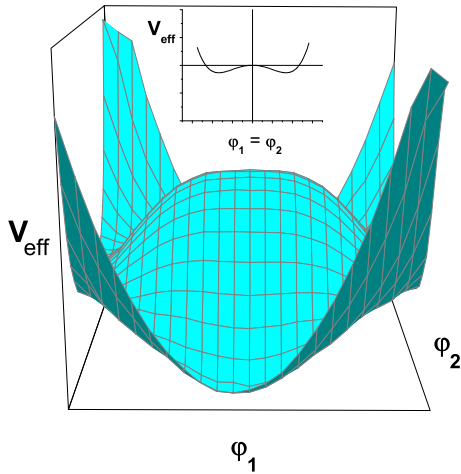} 
\end{center}
\caption{ (Color online) Plot of  the complete effective potential  for $r_{1}$ small, with both couplings finite (schematic), i.e., of the full Eq.~(\ref{eq: num35}). The minima occur for finite values of both order parameters, $\varphi_{1,2}$, such that,  there is  coexistence of phases  (see text). For $\delta_{i}$ finite,  the physical region of the phase diagram is that where $\varphi_{1}$ and $\varphi_{2}$ have the same sign. The plot shows a cut in a 3D graphic for $\varphi_{1}=\varphi_{2}$ (2D graphic) in order to visualize the region of interest. }
\label{Fig.5}
\end{figure}

Finally, we  discuss   the stability of the coexistence phase described by the  complete effective potential in the presence of  both kinds of couplings, i.e., the full Eq.~(\ref{eq: num35}). As shown before, at the classical level  coexistence is always possible whenever $\delta_{1,2} < 0$, even when $r_{2} = 0$, since this lowers the total  \textit{condensation energy}. This is also valid with quantum corrections.
The complete effective potential  assumes a \textit{Mexican hat} shape as shown in Fig.~(\ref{Fig.5}). The minima occur for finite values of both order parameters, $\varphi_{1,2}$,  and consequently there is  a coexistence of phases. Nevertheless, as discussed previously, the $\lambda_{12}$ parameter couples with $\varphi_{1}^{2}$ term and we have competition between the different orderings depending on the quantities $r_{2}$ and $\lambda_{12}$. In addition  the biquadratic and bilinear couplings  also compete producing different trends for the coexistence of phases in the global phase diagram. 

\section{Conclusions \label{sec6}} 

We have investigated in this work the effect of quantum corrections to the phase diagram of systems with competing order parameters.
We  considered the case of a bicritical point in the classical phase diagram,  where two phases vanish simultaneously and continuously, and  also the region of this diagram where there is coexistence between the two phases. 

We studied these problems  in the presence of two types of couplings between the order parameters. One is a  conventional quartic coupling with a positive sign that describes competition between the different orderings~\cite{7,8,9,10}.
The other is a bilinear interaction that is allowed only in special cases where the order parameters have the same symmetry~\cite{15,16}.  These couplings have different effects in the phase diagram.
 
We have obtained that a classical bicritical point is stable to quartic interactions even when quantum fluctuations are taken into account. However, this is not the case in the presence of a bilinear coupling. We have shown that any finite positive bilinear interaction breaks the symmetry of the bicritical point and gives rise to  phase coexistence. This is a purely quantum effect and resembles the physics of the Coleman-Weinberg mechanism~\cite{18} where coupling to a gauge field gives rise to symmetry breaking.

In the region of coexistence, we obtained that the effect of quantum fluctuations  in the presence of quartic interactions 
is to reduce the region of phase  coexistence in the phase diagram. While in the classical case coexistence is allowed in the presence of a quartic coupling,  if the  condition  $\lambda_{12}^2<4\lambda_1 \lambda_2$ is satisfied, when quantum corrections are included no coexistence is possible in the presence of these quartic interactions. This is not the case for a bilinear coupling that favors coexistence even when quantum fluctuations are included.
When both interactions are present, in fact for any finite  bilinear coupling, the different phases can coexist at zero temperature.

For completeness, we point out that the problem where different phases are in close proximity, but do not coexist, has been studied before~\cite{1,1.2,2}. There, critical fluctuations of one phase  interfere with those of the other changing the nature of the quantum phase transitions from continuous to first order~\cite{1,1.2,2}.  

In this paper we describe the dynamics of the different orderings by propagators associated with the same dynamic exponent $z=1$~\cite{2,Mineevz1, Hertz, Mineevz2}.  Preliminary results indicate that the choice of different dynamics and  the dimensionality of the system can affect the final results~\cite{futur}. 

The problem investigated in this paper is  extremely relevant for many systems in condensed matter physics. They range from high temperature superconductors where different phases compete inside the superconducting dome to heavy fermion materials~\cite{superaf,Tuson,Pagliuso,Chen,Sidorov,Chen2,Shermadini}. In the latter antiferromagnetism and superconductivity have clearly been observed in coexistence close to an antiferromagnetic quantum critical point. 
Our results can be useful to identify the form of the interactions between the order parameters in this system. 

\section{ACKNOWLEDGMENTS}

We would like to thank the Brazilian Agencies, CNPq, CAPES and FAPERJ.

\section{Appendix \label{secAP}}
\subsection{Calculation of the integrals of Eq.~(\ref{eq: num14}) \label{integrals}}
In this section we will perform the calculus of the integrals from Eq.~(\ref{eq: num14}) above in text, which is given by,
\begin{eqnarray}
\Gamma^{(1)}&=&\frac{1}{2}\!\int\!\!\!\frac{d^{4}k}{(2\pi)^{4}}\! \ln\left[\left(1+\frac{b_{1}}{k^{2}+r_{1}}\right)\left(1+\frac{b_{2}}{k^{2}+r_{2}}\right)\right] +  \nonumber \\ 
&-&\frac{\left(3\delta_{1}\varphi_{1}^{2}+3\delta_{2}\varphi_{2}^{2}+4\lambda_{12}\varphi_{1}\varphi_{2}\right)^{2}}{2(B^{2}-A^{2})} \times \nonumber \\ 
&\times&\!\int\!\!\!\frac{d^{4}k}{(2\pi)^{4}}\! \left(\frac{1}{k^{2}+A^{2}}-\frac{1}{k^{2}+B^{2}}\right) + ct,
\label{eq: num14.1}
\end{eqnarray}
where,
\begin{eqnarray}
b_{1}&=&12\lambda_{1}\varphi_{1}^{2}+2\lambda_{12}\varphi_{2}^{2}+6\delta_{1}\varphi_{1}\varphi_{2} ; \ A^{2}= r_{1}+b_{1} \nonumber \\
b_{2}&=&12\lambda_{2}\varphi_{2}^{2}+2\lambda_{12}\varphi_{1}^{2}+6\delta_{2}\varphi_{1}\varphi_{2} ; \ B^{2}= r_{2}+b_{2}
\label{eq: num15.1}
\end{eqnarray}

We can calculate the integrals in Eq.~(\ref{eq: num14}) noticing that $d^{d}k=S_{d}k^{d-1}dk$, where $S_{d}=(2 \pi)^{d/2}/\Gamma(d/2)$ and $S_{4}=2 \pi^{2}$. Therefore, the integral in the first line of Eq.~(\ref{eq: num14.1}) is easily calculated~\cite{1.2,2} and yields,
$$I_a=I_1+I_2$$ where,
\begin{eqnarray}
I_{1,2}&=&\frac{\pi^{2}}{(2 \pi)^{4}}\bigg[\frac{\Lambda^{2}}{2}b_{1,2}+\frac{(b_{1,2}+r_{1,2})^{2}}{4}\ln\left(\frac{b_{1,2}+r_{1,2}}{\Lambda^{2}}\right)+ \nonumber \\
&-&\frac{b_{1,2}^{2}}{8}-\frac{r_{1,2}^{2}}{4}\ln\left(\frac{r_{1,2}}{\Lambda^{2}}\right)\bigg]
\label{eq: num16.1}
\end{eqnarray}
with $\Lambda$  an ultraviolet  \textit{cut-off}.

The  integral in the second line of Eq.~(\ref{eq: num14}) has the following solution,
$$I_b=I_A + I_B,$$
where,
\begin{eqnarray}
I_{A,B}&=&\frac{\pi^{2}}{(2 \pi)^{4}}\bigg[\frac{1}{2}\Lambda^{2}+\frac{(A,B)^{2}}{2}\ln\left((A,B)^{2}\right)+ \nonumber \\
&-&\frac{(A,B)^{2}}{2}\ln\left((A,B)^{2}+\Lambda^{2}\right)\bigg].
\label{eq: num17.1}
\end{eqnarray}
The quantities $A$ and $B$ are defined in Eq.~(\ref{eq: num15}).

\subsection{Renormalization \label{secA}}
In this section we will apply the renormalization process to renormalize the effective potential in the presence of both, bilinear and biquadratic couplings in the region of coexistence.  We can get the particular cases easily from this calculation. Expanding the effective potential, Eq.~(\ref{eq: num32}), for  $r_{1}$ small yields,
\begin{eqnarray}
\Gamma^{(1)}&=&\frac{\pi^{2}}{(2 \pi)^{4}}\bigg[\frac{\Lambda^{2}}{2}(b_{1}+b_{2})+\frac{b_{1}^{2}}{4}\ln\left(\frac{b_{1}}{\Lambda^{2}}\right)-\frac{(b_{1}^{2}+b_{2}^{2})}{8}+ \nonumber \\
&+&\frac{(r_{2}+b_{2})^{2}}{4}\ln\left(\frac{r_{2}+b_{2}}{\Lambda^{2}}\right)-\frac{r_{2}^{2}}{4}\ln\left(\frac{r_{2}}{\Lambda^{2}}\right)+ \nonumber \\
&+&\frac{\left(3\delta_{1}\varphi_{1}^{2}+3\delta_{2}\varphi_{2}^{2}+4\lambda_{12}\varphi_{1}\varphi_{2}\right)^{2}}{4}\!\!\times \nonumber \\
&\times&\left(\ln\left(\frac{1}{\Lambda^{2}}\right)+\frac{B^{2}\ln(B^{2})-A^{2}\ln(A^{2})}{B^{2}-A^{2}}\right)\!\!\bigg] + ct
\label{eq: num38}
\end{eqnarray}

The \textit{counterterms} necessary for renormalization with additional quantum corrections must have terms of the same form as those from the classical free energy, Eq.~(\ref{eq: num130}). Thus, we have to calculate terms of the type,
\begin{eqnarray}
\underbrace{\frac{1}{2}C_{1}\varphi_{1}^{2} \ , \ \frac{1}{2}C_{2}\varphi_{2}^{2} \ , \ C_{3}\varphi_{1}\varphi_{2}}_{quadratic \ terms} \ , \ \frac{1}{4!}D_{1}\varphi_{1}^{4} \ , \nonumber \\
\frac{1}{4!}D_{2}\varphi_{2}^{4} \ , \ \frac{1}{4}D_{3}\varphi_{1}^{2}\varphi_{2}^{2} \ , \ \frac{1}{6}D_{4}\varphi_{1}^{3}\varphi_{2} \, \ \frac{1}{6}D_{5}\varphi_{1}\varphi_{2}^{3};
\label{eq: num39}
\end{eqnarray}
The calculation of these terms requires that we define, as usual~\cite{1.2,2,17,18}, the following conditions,
\begin{eqnarray}
\left[\frac{d^{2} \Gamma^{(1)}}{d \varphi_{1}^{2}}\right]_{\varphi_{1,2}=0}= r_{1} = 0 ; \left[\frac{d^{4} \Gamma^{(1)}}{d \varphi_{1}^{4}}\right]_{\varphi_{1}=\left\langle \varphi_{1}\right\rangle,\varphi_{2}=0} = \lambda_{1} \nonumber \\
\left[\frac{d^{2} \Gamma^{(1)}}{d \varphi_{2}^{2}}\right]_{\varphi_{1,2}=0}= r_{2}  ; \left[\frac{d^{4} \Gamma^{(1)}}{d \varphi_{2}^{4}}\right]_{\varphi_{1}=0,\varphi_{2}=\left\langle \varphi_{2}\right\rangle} = \lambda_{2} \nonumber \\
\left[\frac{d^{4} \Gamma^{(1)}}{d \varphi_{1}^{2}\varphi_{2}^{2}}\right]_{\varphi_{1,2}=\left\langle \varphi_{1,2}\right\rangle}= \lambda_{12} ; \left[\frac{d^{2} \Gamma^{(1)}}{d \varphi_{1}\varphi_{2}}\right]_{\varphi_{1,2}=\left\langle \varphi_{1,2}\right\rangle}= \gamma \nonumber \\
\left[\frac{d^{4} \Gamma^{(1)}}{d \varphi_{1}^{3}\varphi_{2}}\right]_{\varphi_{1,2}=\left\langle \varphi_{1,2}\right\rangle}= \delta_{1} ; \left[\frac{d^{2} \Gamma^{(1)}}{d \varphi_{1}\varphi_{2}^{3}}\right]_{\varphi_{1,2}=\left\langle \varphi_{1,2}\right\rangle}= \delta_{2}\nonumber \\
\label{eq: num40}
\end{eqnarray}
where $\left\langle \varphi_{1,2}\right\rangle$ correspond to the minima of the effective potential.
Proceeding with the calculation of the derivatives in Eq.~(\ref{eq: num40}), we obtain the \textit{counterterms},
\begin{eqnarray}
\frac{1}{2}C_{1}\varphi_{1}^{2}&=& - \frac{1}{2}\frac{\pi^{2}}{\left(2 \pi\right)^{4}}\bigg[\Lambda^{2}(12\lambda_{1}+2\lambda_{12})+\nonumber \\
&+&r_{2}\lambda_{12}\left(1+2\ln\left(\frac{|r_{2}|}{\Lambda^{2}}\right)\right)\bigg]\varphi_{1}^{2}
\label{eq: num41-Ap}
\end{eqnarray}
\begin{eqnarray}
\frac{1}{2}C_{2}\phi_{2}^{2}&=& - \frac{1}{2}\frac{\pi^{2}}{\left(2 \pi\right)^{4}}\bigg[\Lambda^{2}(12\lambda_{2}+2\lambda_{12})+\nonumber \\
&+&6r_{2}\lambda_{2}\left(1+2\ln\left(\frac{|r_{2}|}{\Lambda^{2}}\right)\right)\bigg]\varphi_{2}^{2}
\label{eq: num42}
\end{eqnarray}
\begin{eqnarray}
C_{3}\varphi_{1}\varphi_{2}&=& - \frac{\pi^{2}}{\left(2 \pi\right)^{4}}\bigg[\Lambda^{2}(3\delta_{1}+3\delta_{2})+\nonumber \\
&+&\frac{3}{2}r_{2}\delta_{2}\left(1+2\ln\left(\frac{|r_{2}|}{\Lambda^{2}}\right)\right)\bigg]\varphi_{1}\varphi_{2}
\label{eq: num42.2}
\end{eqnarray}
\begin{eqnarray}
\frac{1}{4!}D_{1}\phi_{1}^{4}&=& - \frac{1}{4!}\frac{\pi^{2}}{\left(2 \pi\right)^{4}}\bigg[16\lambda_{1}^{2}\bigg(198+\nonumber \\
&+&54 \ln\left(\frac{12\lambda_{1} \left\langle \varphi_{1}\right\rangle^{2}}{\Lambda^{2}}\right)\bigg)+ \nonumber \\
&+&8\lambda_{12}^{2}\left(11+3 \ln\left(\frac{2\lambda_{12}\left\langle \varphi_{1}\right\rangle^{2}}{\Lambda^{2}}\right)\right)+\nonumber \\
&+&54\delta_{1}^{2} \ln\left(\frac{1}{\Lambda^{2}}\right)\bigg]\varphi_{1}^{4}
\label{eq: num43}
\end{eqnarray}
\begin{eqnarray}
\frac{1}{4!}D_{2}\varphi_{2}^{4}&=&- \frac{1}{4!}\frac{\pi^{2}}{\left(2 \pi\right)^{4}}\bigg[16\lambda_{2}^{2}\bigg(198+\nonumber \\
&+&54\ln\left(\frac{12\lambda_{2}\left\langle \varphi_{2}\right\rangle^{2}}{\Lambda^{2}}\right)\bigg)+ \nonumber \\
&+&8\lambda_{12}^{2}\left(11+3 \ln\left(\frac{2\lambda_{12} \left\langle \varphi_{2}\right\rangle^{2}}{\Lambda^{2}}\right)\right)+\nonumber \\
&+&54\delta_{2}^{2} \ln\left(\frac{1}{\Lambda^{2}}\right)\bigg]\varphi_{2}^{4}
\label{eq: num44}
\end{eqnarray}
\begin{eqnarray}
\frac{1}{4}D_{3}\phi_{1}^{2}\phi_{2}^{2}&=& - \frac{1}{4}\frac{\pi^{2}}{\left(2 \pi\right)^{4}}\bigg[\delta_{1}^{2}\bigg[90+ \nonumber \\
&+&36 \ln\left(\frac{6\delta_{1} \left\langle \varphi_{1}\right\rangle\left\langle \varphi_{2}\right\rangle }{\Lambda^{2}}\right)\bigg]+\nonumber \\
&+&16\lambda_{1}\lambda_{12}\left(9+3 \ln\left(\frac{12\lambda_{1} \left\langle \varphi_{1}\right\rangle^{2}}{\Lambda^{2}}\right)\right)+\nonumber \\
&+&\delta_{2}^{2}\left(90+36 \ln\left(\frac{6\delta_{2} \left\langle \varphi_{1}\right\rangle\left\langle \varphi_{2}\right\rangle}{\Lambda^{2}}\right)\right)+ \nonumber \\
&+&16\lambda_{2}\lambda_{12}\left(9+3 \ln\left(\frac{12\lambda_{2} \left\langle \varphi_{2}\right\rangle^{2}}{\Lambda^{2}}\right)\right)+\nonumber \\
&+&18\delta_{1}\delta_{2}\ln\left(\frac{1}{\Lambda^{2}}\right)+\nonumber \\
&+&16\lambda_{12}^{2}\ln\left(\frac{1}{\Lambda^{2}}\right)\bigg]\varphi_{1}^{2}\varphi_{2}^{2}
\label{eq: num45}
\end{eqnarray}
\begin{eqnarray}
\frac{1}{6} D_{4} \phi_{1}^{3} \phi_{2}&=& - \frac{1}{6}\frac{\pi^{2}}{\left( 2 \pi \right)^{4}} \bigg[ 8 \lambda_{1}\delta_{1} \bigg[ 99+\nonumber \\
&+&27 \ln \left( \frac{12\lambda_{1} \left\langle \varphi_{1}\right\rangle^{2}}{\Lambda^{2}} \right)\bigg] + \nonumber \\
&+&4\delta_{2}\lambda_{12}\left(27+9 \ln\left(\frac{2\lambda_{12} \left\langle \varphi_{2}\right\rangle^{2}}{\Lambda^{2}}\right)\right)+\nonumber \\
&+&36\delta_{1}\lambda_{12}\ln\left(\frac{1}{\Lambda^{2}}\right) \bigg] \varphi_{1}^{3}\varphi_{2}
\label{eq: num46}
\end{eqnarray}
\begin{eqnarray}
\frac{1}{6}D_{5}\phi_{1}\phi_{2}^{3}&=& - \frac{1}{6}\frac{\pi^{2}}{\left(2 \pi\right)^{4}}\bigg[8\lambda_{2}\delta_{2}\bigg[99+\nonumber \\
&+&27 \ln\left(\frac{12\lambda_{2} \left\langle \varphi_{2}\right\rangle^{2} }{\Lambda^{2}}\right)\bigg]+ \nonumber \\
&+&4\delta_{1}\lambda_{12}\left(27+9 \ln\left(\frac{2\lambda_{12} \left\langle \varphi_{1}\right\rangle^{2}}{\Lambda^{2}}\right)\right)+\nonumber \\
&+&36\delta_{2}\lambda_{12}\ln\left(\frac{1}{\Lambda^{2}}\right)\bigg]\varphi_{1}\varphi_{2}^{3}
\label{eq: num46.1}
\end{eqnarray}
Replacing the \textit{counterterms} into Eq.~(\ref{eq: num38}) we can write the renormalized effective potential in the form,
\begin{eqnarray}
&&   \;\;\;\; V_{eff}= r_{2}\varphi_{2}^{2}+\lambda_{1}\varphi_{1}^{4}+\lambda_{2}\varphi_{2}^{4} +  \nonumber \\
&&   \;\;\;\;\;\;\;\;\;\;\;\; +\lambda_{12} \varphi_{1}^{2} \varphi_{2}^{2}+\delta_{1} \varphi_{1}^{3}\varphi_{2}+\delta_{2} \varphi_{1}\varphi_{2}^{3}+ 
\nonumber \\
&&   \;\;\;\;\;\;\;\;\;\;\;\; +\frac{\pi^{2}}{(2 \pi)^{4}}\bigg[\bigg[36\lambda_{1}^{2}\ln\left(\frac{b_{1}}{12\lambda_{1}\left\langle \phi_{1}\right\rangle^{2}}\right)+ \nonumber \\
&&   \;\;\;\;\;\;\;\;\;\;\;\; +\lambda_{12}^{2}\ln\left(\frac{|r_{2}+b_{2}|}{2\lambda_{12}\left\langle \varphi_{1}\right\rangle^{2}}\right)\bigg]\varphi_{1}^{4}+ \nonumber \\
&&   \;\;\;\;\;\;\;\;\;\;\;\; +\bigg[36\lambda_{2}^{2}\ln\left(\frac{|r_{2}+b_{2}|}{12\lambda_{2}\left\langle \varphi_{2}\right\rangle^{2}}\right)+  \nonumber \\
&&   \;\;\;\;\;\;\;\;\;\;\;\; +\lambda_{12}^{2}\ln\left(\frac{b_{1}}{2\lambda_{12}\left\langle \phi_{2}\right\rangle^{2}}\right)\bigg]\varphi_{2}^{4}+ \nonumber \\
&&   \;\;\;\;\;\;\;\;\;\;\;\; +\bigg[9\delta_{1}^{2}\ln\left(\frac{b_{1}}{6 \delta_{1}\left\langle \varphi_{1}\right\rangle\left\langle \varphi_{2}\right\rangle}\right)+  \nonumber \\
&&   \;\;\;\;\;\;\;\;\;\;\;\; +12\lambda_{1}\lambda_{12}\ln\left(\frac{b_{1}}{12 \lambda_{1}\left\langle \varphi_{1}\right\rangle^{2}}\right)+ \nonumber \\
&&   \;\;\;\;\;\;\;\;\;\;\;\; +9\delta_{2}^{2}\ln\left(\frac{|r_{2}+b_{2}|}{6 \delta_{2}\left\langle \varphi_{1}\right\rangle\left\langle \varphi_{2}\right\rangle}\right)+  \nonumber \\
&&   \;\;\;\;\;\;\;\;\;\;\;\; +12\lambda_{2}\lambda_{12}\ln\left(\frac{|r_{2}+b_{2}|}{12 \lambda_{2}\left\langle \varphi_{2}\right\rangle^{2}}\right)\bigg]\varphi_{1}^{2}\varphi_{2}^{2}+ \nonumber \\
&&   \;\;\;\;\;\;\;\;\;\;\;\; +\bigg[36\delta_{1}\lambda_{1}\ln\left(\frac{b_{1}}{12\lambda_{1}\left\langle \varphi_{1}\right\rangle^{2}}\right)+  \nonumber \\
&&   \;\;\;\;\;\;\;\;\;\;\;\; +6\delta_{2}\lambda_{12}\ln\left(\frac{|r_{2}+b_{2}|}{2\lambda_{12}\left\langle \varphi_{2}\right\rangle^{2}}\right)\bigg]\varphi_{1}^{3}\varphi_{2}+ \nonumber \\
&&   \;\;\;\;\;\;\;\;\;\;\;\; +\bigg[36\delta_{2}\lambda_{2}\ln\left(\frac{|r_{2}+b_{2}|}{12\lambda_{2}\left\langle \varphi_{2}\right\rangle^{2}}\right)+  \nonumber \\
&&   \;\;\;\;\;\;\;\;\;\;\;\; +6\delta_{1}\lambda_{12}\ln\left(\frac{b_{1}}{2\lambda_{12}\left\langle \varphi_{1}\right\rangle^{2}}\right)\bigg]\varphi_{1}\varphi_{2}^{3}+ \nonumber \\
&+&\left(3r_{2}\delta_{2}\varphi_{1}\varphi_{2}+6r_{2}\lambda_{2}\varphi_{2}^{2}+r_{2}\lambda_{12}\varphi_{1}^{2}\right)\ln\left(\frac{|r_{2}+b_{2}|}{|r_{2}|}\right)+ \nonumber \\
&-&\bigg(150\lambda_{1}^{2}\varphi_{1}^{4}+\frac{25}{6}\lambda_{12}^{2}\varphi_{1}^{4}+150\lambda_{2}^{2}\varphi_{2}^{4}+\frac{25}{6}\lambda_{12}^{2}\varphi_{2}^{4}+ \nonumber \\
&&   \;\;\;\;\;\;\;\;\;\;\;\; +27\delta_{1}^{2}\varphi_{1}^{2}\varphi_{2}^{2}+27\delta_{2}^{2}\varphi_{1}^{2}\varphi_{2}^{2}+ \nonumber \\
&&   \;\;\;\; +42\lambda_{1}\lambda_{12}\varphi_{1}^{2}\varphi_{2}^{2}+42\lambda_{2}\lambda_{12}\varphi_{1}^{2}\varphi_{2}^{2}+150\delta_{1}\lambda_{1}\varphi_{1}^{3}\varphi_{2}+ \nonumber \\
&&   \;\;\;\; +21\delta_{2}\lambda_{12}\varphi_{1}^{3}\varphi_{2}+150\delta_{2}\lambda_{2}\varphi_{1}\varphi_{2}^{3}+21\delta_{1}\lambda_{12}\varphi_{1}\varphi_{2}^{3}+ \nonumber \\
&&   \;\;\;\;\;\;\;\;\;\;\;\; +\frac{3}{2}r_{2}\delta_{2}\varphi_{1}\varphi_{2}+3r_{2}\lambda_{2}\varphi_{2}^{2}+\frac{1}{2}r_{2}\lambda_{12}\varphi_{1}^{2}\bigg)+ \nonumber \\
&&   \;\;\;\;\;\;\;\;\;\;\;\; +\frac{(3\delta_{1}\varphi_{1}^{2}+3\delta_{2}\varphi_{2}^{2}+4\lambda_{12}\varphi_{1}\varphi_{2})^{2}}{4}\times  \nonumber \\
&&   \;\;\;\;\;\;\;\;\;\;\;\; \times\left(\frac{B^{2}\ln(B^{2})-b_{1}\ln(b_{1})}{B^{2}-b_{1}}\right)\bigg]
\label{eq: num46-Ap2}
\end{eqnarray}
where,
\begin{eqnarray}
&&   \;\;\;\; b_{1}=12\lambda_{1}\varphi_{1}^{2}+2\lambda_{12}\varphi_{2}^{2}+6\delta_{1}\varphi_{1}\varphi_{2} \nonumber \\
b_{2}&=&12\lambda_{2}\varphi_{2}^{2}+2\lambda_{12}\varphi_{1}^{2}+6\delta_{2}\varphi_{1}\varphi_{2} ; B^{2}=|r_{2}+b_{2}|
\label{eq: num47}
\end{eqnarray}

\subsection{Effective couplings parameters renormalized - $r_{1}$ and $r_{2}$ expansion \label{secB}}

In this section, we give details of the calculation of the renormalized effective couplings parameters and the higher order field terms for the  bicritical point case and  that leads to Eq.~(\ref{eq: num19}) in text.   The calculation of the effective potential for this case proceeds  in the same way as in the previous section, but expanding both in $r_{1}$ and $r_{2}$ concomitantly, i.e., starting from Eq.~(\ref{eq: num18}). Thus, following the procedure of the previous section step by step we get the renormalized effective couplings parameters  in the form,
\begin{eqnarray}
\tilde{\lambda}_{1}&=& \lambda_{1} + \frac{\pi^{2}}{(2 \pi)^{4}}\bigg[36\lambda_{1}^{2}\left(\ln\left(\frac{b_{1}}{12\lambda_{1}\left\langle \varphi_{1}\right\rangle^{2}}\right)-\frac{25}{6}\right) +\nonumber \\
&+&\lambda_{12}^{2}\left(\ln\left(\frac{b_{2}}{2\lambda_{12}\left\langle \varphi_{1}\right\rangle^{2}}\right)-\frac{25}{6}\right)+\nonumber \\
&+&\frac{9}{4}\delta_{1}^{2}\left(\ln\left(\frac{1}{6\delta_{1}\left\langle \varphi_{1}\right\rangle\left\langle \varphi_{2}\right\rangle}\right)-\frac{25}{6}\right)\bigg]
\label{eq: num48}
\end{eqnarray}
\begin{eqnarray}
\tilde{\lambda}_{2}&=& \lambda_{2} + \frac{\pi^{2}}{(2 \pi)^{4}}\bigg[36\lambda_{2}^{2}\left(\ln\left(\frac{b_{2}}{12\lambda_{2}\left\langle \varphi_{2}\right\rangle^{2}}\right)-\frac{25}{6}\right)+\nonumber \\
&+&\lambda_{12}^{2}\left(\ln\left(\frac{b_{1}}{2\lambda_{12}\left\langle \varphi_{2}\right\rangle^{2}}\right)-\frac{25}{6}\right)+\nonumber \\
&+&\frac{9}{4}\delta_{2}^{2}\left(\ln\left(\frac{1}{6\delta_{2}\left\langle \varphi_{1}\right\rangle\left\langle \varphi_{2}\right\rangle}\right)-\frac{25}{6}\right)\bigg]
\label{eq: num49}
\end{eqnarray}
\begin{eqnarray}
\tilde{\lambda}_{12}&=& \lambda_{12} + \frac{\pi^{2}}{(2 \pi)^{4}}\bigg[9\delta_{1}^{2}\left(\ln\left(\frac{b_{1}}{6 \delta_{1}\left\langle \phi_{1}\right\rangle\left\langle \phi_{1}\right\rangle}\right)-3\right)+\nonumber \\
&+&9\delta_{2}^{2}\left(\ln\left(\frac{b_{2}}{6 \delta_{2}\left\langle \phi_{1}\right\rangle\left\langle \phi_{2}\right\rangle}\right)-3\right)+\nonumber \\
&+&12\lambda_{1}\lambda_{12}\left(\ln\left(\frac{b_{1}}{12 \lambda_{1}\left\langle \phi_{1}\right\rangle^{2}}\right)-\frac{7}{2}\right)+\nonumber \\
&+&12\lambda_{2}\lambda_{12}\left(\ln\left(\frac{b_{2}}{12 \lambda_{2}\left\langle \phi_{2}\right\rangle^{2}}\right)-\frac{7}{2}\right)\bigg]
\label{eq: num50}
\end{eqnarray}
\begin{eqnarray}
\tilde{\delta_{1}}&=& \delta_{1} + \frac{\pi^{2}}{(2 \pi)^{4}}\bigg[36\delta_{1}\lambda_{1}\left(\ln\left(\frac{b_{1}}{12\lambda_{1}\left\langle \varphi_{1}\right\rangle^{2}}\right)-\frac{25}{6}\right)+\nonumber \\
&+&6\delta_{2}\lambda_{12}\left(\ln\left(\frac{b_{2}}{2\lambda_{12}\left\langle \varphi_{2}\right\rangle^{2}}\right)-\frac{7}{2}\right)\bigg]
\label{eq: num51}
\end{eqnarray}
\begin{eqnarray}
\tilde{\delta_{2}}&=& \delta_{2} + \frac{\pi^{2}}{(2 \pi)^{4}}\bigg[36\delta_{2}\lambda_{2}\left(\ln\left(\frac{b_{2}}{12\lambda_{2}\left\langle \varphi_{2}\right\rangle^{2}}\right)-\frac{25}{6}\right)+\nonumber \\
&+&6\delta_{1}\lambda_{12}\left(\ln\left(\frac{b_{1}}{2\lambda_{12}\left\langle \varphi_{1}\right\rangle^{2}}\right)-\frac{7}{2}\right)\bigg]
\label{eq: num51.1}
\end{eqnarray}
\begin{eqnarray}
O(\varphi_{1,2}^{6})&=&\frac{\pi^{2}}{(2 \pi)^{4}}\bigg[\frac{\left(3\delta_{1}\varphi_{1}^{2}+3\delta_{2}\varphi_{2}^{2}+4\lambda_{12}\varphi_{1}\varphi_{2}\right)^{2}}{4}\times \nonumber \\
&\times&\left(\frac{b_{1}\ln (b_{1})-b_{2}\ln (b_{2}) }{b_{1}-b_{2}}\right)\bigg]
\label{eq: num51.2}
\end{eqnarray}
where,
\begin{eqnarray}
b_{1}&=&12\lambda_{1}\varphi_{1}^{2}+2\lambda_{12}\varphi_{2}^{2}+6\delta_{1}\varphi_{1}\varphi_{2} ; \nonumber \\
b_{2}&=&12\lambda_{2}\varphi_{2}^{2}+2\lambda_{12}\varphi_{1}^{2}+6\delta_{2}\varphi_{1}\varphi_{2} ; 
\label{eq: num52}
\end{eqnarray}

\subsection{Effective couplings parameters renormalized - $r_{1}$ expansion \label{secC}}

Analogously to the  previous section of this Appendix, after obtaining the effective potential, we can write the  renormalized effective coupling parameters and the higher order field terms in the coexistence phase, Eq.~(\ref{eq: num35}), as,
\begin{eqnarray}
\lambda_{1}'&=& \lambda_{1} + \frac{\pi^{2}}{(2 \pi)^{4}}\bigg[36\lambda_{1}^{2}\left(\ln\left(\frac{b_{1}}{12\lambda_{1}\left\langle \phi_{1}\right\rangle^{2}}\right)-\frac{25}{6}\right)+\nonumber \\
&+&\lambda_{12}^{2}\left(\ln\left(\frac{|r_{2}+b_{2}|}{2\lambda_{12}\left\langle \varphi_{1}\right\rangle^{2}}\right)-\frac{25}{6}\right)\bigg]
\label{eq: num54}
\end{eqnarray}
\begin{eqnarray}
\lambda_{2}'&=& \lambda_{2} + \frac{\pi^{2}}{(2 \pi)^{4}}\bigg[36\lambda_{2}^{2}\left(\ln\left(\frac{|r_{2}+b_{2}|}{12\lambda_{2}\left\langle \varphi_{2}\right\rangle^{2}}\right)-\frac{25}{6}\right)+\nonumber \\
&+&\lambda_{12}^{2}\left(\ln\left(\frac{b_{1}}{2\lambda_{12}\left\langle \phi_{2}\right\rangle^{2}}\right)-\frac{25}{6}\right)\bigg]
\label{eq: num55}
\end{eqnarray}
\begin{eqnarray}
\lambda_{12}'&=& \lambda_{12} + \frac{\pi^{2}}{(2 \pi)^{4}}\bigg[9\delta_{1}^{2}\left(\ln\left(\frac{b_{1}}{6 \delta_{1}\left\langle \phi_{1}\right\rangle\left\langle \phi_{2}\right\rangle}\right)-3\right)+\nonumber \\
&+&9\delta_{2}^{2}\left(\ln\left(\frac{|r_{2}+b_{2}|}{6\delta_{2}\left\langle \varphi_{1}\right\rangle\left\langle \varphi_{2}\right\rangle}\right)-3\right)+\nonumber \\
&+&12\lambda_{1}\lambda_{12}\left(\ln\left(\frac{b_{1}}{12 \lambda_{1}\left\langle \phi_{1}\right\rangle^{2}}\right)-\frac{7}{2}\right)+\nonumber \\
&+&12\lambda_{2}\lambda_{12}\left(\ln\left(\frac{|r_{2}+b_{2}|}{12 \lambda_{2}\left\langle \phi_{2}\right\rangle^{2}}\right)-\frac{7}{2}\right)\bigg]
\label{eq: num56}
\end{eqnarray}
\begin{eqnarray}
\delta_{1}'&=& \delta_{1}+ \frac{\pi^{2}}{(2 \pi)^{4}}\bigg[36\delta_{1}\lambda_{1}\left(\ln\left(\frac{b_{1}}{12\lambda_{1}\left\langle \varphi_{1}\right\rangle^{2}}\right)-\frac{25}{6}\right)+\nonumber \\
&+&6\delta_{2}\lambda_{12}\left(\ln\left(\frac{|r_{2}+b_{2}|}{2\lambda_{12}\left\langle \varphi_{2}\right\rangle^{2}}\right)-\frac{7}{2}\right)\bigg]
\label{eq: num57}
\end{eqnarray}
\begin{eqnarray}
\delta_{2}'&=& \delta_{2}+ \frac{\pi^{2}}{(2 \pi)^{4}}\bigg[36\delta_{2}\lambda_{2}\left(\ln\left(\frac{|r_{2}+b_{2}|}{12\lambda_{2}\left\langle \varphi_{2}\right\rangle^{2}}\right)-\frac{25}{6}\right)+\nonumber \\
&+&6\delta_{1}\lambda_{12}\left(\ln\left(\frac{b_{1}}{2\lambda_{12}\left\langle \varphi_{1}\right\rangle^{2}}\right)-\frac{7}{2}\right) \bigg]
\label{eq: num57.1}
\end{eqnarray}
\begin{eqnarray}
\tilde{\rho}(\varphi_{1,2})&=&\frac{\pi^{2}}{(2 \pi)^{4}}\bigg[\frac{\left(3\delta_{1}\varphi_{1}^{2}+3\delta_{2}\varphi_{2}^{2}+4\lambda_{12}\varphi_{1}\varphi_{2}\right)^{2}}{4}\times \nonumber \\
&\times& \left(\frac{B^{2}\ln (B^{2})-b_{1}\ln (b_{1}) }{B^{2}-b_{1}}\right)\bigg]
\label{eq: num57.2}
\end{eqnarray}
where,
\begin{eqnarray}
b_{1}&=&12\lambda_{1}\varphi_{1}^{2}+2\lambda_{12}\varphi_{2}^{2}+6\delta_{1}\varphi_{1}\varphi_{2}; \nonumber \\
b_{2}&=&12\lambda_{2}\varphi_{2}^{2}+2\lambda_{12}\varphi_{1}^{2}+6\delta_{2}\varphi_{1}\varphi_{2}; B^{2}=|r_{2}+b_{2}|
\label{eq: num58}
\end{eqnarray}


\begin{thebibliography}{99}

\bibitem{cdw} Thomas Gruner, Dongjin Jang, Zita Huesges, Raul Cardoso-Gil, Gerhard H. Fecher, Michael M. Koza, Oliver Stockert, Andrew P. Mackenzie, Manuel Brando and Christoph Geibel, Nat. Phys.{\bf 13}, 967 (2017).
\bibitem{0} Frank Steglich and Steffen Wirth, Rep. Prog. Phys. {\bf 79} 084502  (2016); P. G. Pagliuso, C. Petrovic, R. Movshovich, D. Hall, M. F. Hundley, J. L. Sarrao, J. D. Thompson, and Z. Fisk, Phys. Rev. {\bf B 64}, 100503(R) (2001); K. Chen, F. Strigari, M. Sundermann, Z. Hu, Z. Fisk, E. D. Bauer, P. F. S. Rosa, J. L. Sarrao, J. D. Thompson, J. Herrero-Martin, E. Pellegrin, D. Betto, K. Kummer, A. Tanaka, S. Wirth, and A. Severing, Phys. Rev. {\bf B 97}, 045134 (2018).
\bibitem{sachdev} S. Sachdev, \textit{Quantum Phase Transitions}, Cambridge University Press, UK (1999).
\bibitem{1} M.A. Continentino, Brazilian Journal of Physics, vol. 35, no. 1, March, 2005; M.A. Continentino, A.S. Ferreira, J. M. M. M. {\bf 310}  828 (2007).
\bibitem{1.2} M.A. Continentino, {\it Quantum scaling in many-body systems: an approach to quantum phase transitions}, Cambridge University Press, 2017.
\bibitem{1.1} S. Doniach, Physica {\bf B 91}, 231 (1977).
\bibitem{rkky}J. D. Thompson and J. M. Lawrence in \textit{Handbook on the Physics and Chemistry of Rare Earths, Lanthanides/Actinides:Physics-II}, edited by K.A. Gschneider Jr., L. Eyring, G.H. Lander, and G.R. Choppin , Elsevier Science B.V., Chapter 133 (19), 383 (1994).
\bibitem{2} A.S. Ferreira, M.A. Continentino and E.C. Marino, Phys. Rev. B 70, 174507 (2004).
\bibitem{3} H. Luetkens, H.-H. Klauss, M. Kraken, F. J. Litterst, T. Dellmann, R. Klingeler, C. Hess, R. Khasanov, A. Amato, C. Baines,
M. Kosmala, O. J. Schumann, M. Braden, J. Hamann-Borrero, N. Leps, A. Kondrat, G. Behr, J. Werner, and B. Büchner, Nature Mater. 8, 305 (2009).
\bibitem{4} A. J. Drew, Ch. Niedermayer, P. J. Baker, F. L. Pratt, S. J. Blundell, T. Lancaster, R. H. Liu, G. Wu, X. H. Chen, I. Watanabe, V.
K. Malik, A. Dubroka, M. Rössle, K. W. Kim, C. Baines, and C. Bernhard, Nature Mater. 8, 310 (2009).
\bibitem{5} C. R. Rotundu, D. T. Keane, B. Freelon, S. D. Wilson, A. Kim, P. N. Valdivia, E. Bourret-Courchesne, and R. J. Birgeneau, Phys. Rev. B 80, 144517 (2009).
\bibitem{6}T. Goko, A. A. Aczel, E. Baggio-Saitovitch, S. L. Bud'ko, P. C Canfield, J. P. Carlo, G. F. Chen, P. Dai, A. C. Hamann, W. Z. Hu, H. Kageyama, G. M. Luke, J. L. Luo, B. Nachumi, N. Ni, D. Reznik, D. R. Sanchez-Candela, A. T. Savici, K. J. Sikes, N. L. Wang, C. R. Wiebe, T. J. Williams, T. Yamamoto, W. Yu, and Y. J. Uemura, Phys. Rev. B 80, 024508 (2009).
\bibitem{7} R.M. Fernandes and J. Schmalian, Phys. Rev. B 82, 014521 (2010).
\bibitem{belitz} D. Belitz, T. R. Kirkpatrick, and J. Rollb\"uhler, PRL 94, 247205 (2005).
\bibitem{belitz2} D. Belitz, T. R. Kirkpatrick, Thomas Vojta, Reviews of Modern Physics, vol. 77, April (2005).
\bibitem{belitz3} T. Vojta, D. Belitz, T. R. Kirkpatrick, and R. Narayanan, Ann. Phys. (Leipzig) 8 (1999) 7–9, 593 – 602.
\bibitem{larkinpikin} A. L. Larkin and S. A. Pikin, Zh. Eksp. Teor. Fiz. {\bf 56}, 1664 (1969).
\bibitem{belitz4} T. R. Kirkpatrick, D. Belitz, Thomas Vojta, and R. Narayanan, Phys. Rev. Lett. 87, 127003 (2001).
\bibitem{8} P. Santini and G. Amoretti, Phys. Rev. Lett. 73, 1027 (1994).
\bibitem{9} V. Barzykin and L. P. Gor'kov, Phys. Rev. Lett. 70, 2479 (1993).
\bibitem{10} D.G. Barci, R.V. Clarim and  N.L. Silva J\'unior, Phys. Rev. B 94, 184507 (2016).
\bibitem{15} V. P. Mineev and M. E. Zhitomirsky, Phys. Rev. B 72, 014432 (2005).
\bibitem{16} D. F. Agterberg and M. B. Walker, Phys. Rev. B 50, 563 (1994).
\bibitem{11} A. P. Ramirez, P. Coleman, P. Chandra, E. Br\"uck, A. A. Menovsky, Z. Fisk, and E. Bucher, Phys. Rev. Lett. 68, 2680 (1992).
\bibitem{12} H. Ikeda and Y. Ohashi, Phys. Rev. Lett. 81, 3723 (1998).
\bibitem{13} P. Chandra, P. Coleman, J. A. Mydosh, and V. Tripathi, Nature (London) 417, 831 (2002); P. Chandra, P. Coleman, and J. A.
Mydosh, Physica B 312–313, 397 (2002).
\bibitem{14} M. Zacharias, I. Paul and M. Garst, Phys. Rev. Lett. 115, 025703 (2015).
\bibitem{14.1}M. E. Zhitomirsky and V.-H. Dao, Phys. Rev. B 69, 054508 (2004).
\bibitem{17} G. Jona-Lasinio, Nuovo Cimento 34, 1790 (1964).
\bibitem{18} S. Coleman and E. Weinberg, Phys. Rev. D 7, 1888 (1973).
\bibitem{Mineevz1} V.P. Mineev, Pis'ma Zh. \'Eksp. Teor Fiz. 66, 655 (1997)[JETP Lett. 66, 693 (1997)].
\bibitem{Hertz} J.A. Hertz, Phys. Rev. {\bf B  14}, 3 (1976).
\bibitem{Mineevz2} V.P. Mineev, M. Sigrist, Phys. Rev. {\bf B  63}, 172504 (2001).
\bibitem{dalson} D. E. Almeida, R. M. Fernandes, and E. Miranda, Phys. Rev. {\bf B  96}, 014514 (2017).
\bibitem{futur} N. L. Silva J\'unior, D. Barci and M. A. Continentino, in preparation.
\bibitem{superaf} R. Movshovich, T. Graf, D. Mandrus, J. D. Thompson, J. L. Smith, and Z. Fisk, Phys. Rev. {\bf B 53}, 8241 (1996); N.  D. Matur, F. M. Grosche, S. R. Julian, I. R. Walker, D. M. Freye, R. K. W. Haselwimmer and G. G. Lonzarich,  Nature {\bf 394}, 39 (1998).
\bibitem{Tuson} T. Park et al, Nature Letters {\bf 440} (2006).
\bibitem{Pagliuso} P.G. Pagliuso et al, Phys. Rev. B, {\bf 64}, 100503(R) (2001).
\bibitem{Chen} K. Chen et al, Phys. Rev. B, {\bf 97}, 045134 (2018).
\bibitem{Sidorov} V.A. Sidorov et al, Phys. Rev. Lett., {\bf 89}, 157004 (2002).
\bibitem{Chen2} G.F. Chen, K.Matsubayashi, S. Bah, K. Deguchi and N.K. Sato, Phys. Rev. Lett., {\bf 97}, 017005 (2006).
\bibitem{Shermadini} Z. Shermadini, Phys. Rev. Lett., {\bf 106}, 117602 (2011).

\end{thebibliography}
\end{document}